\documentclass[
aps,           
pra,            
twocolumn,      
letterpaper,    
showpacs,       
preprintnumbers,
amsmath,        
amssymb,        
floatfix, 10pt]{revtex4-2}

\usepackage{graphicx}    
\usepackage{color}       
\usepackage{bm}          
\usepackage{braket}      
\usepackage[T1]{fontenc}
\usepackage{dcolumn}
\usepackage{booktabs}
\usepackage{amsmath,amssymb}
\usepackage[caption=false]{subfig}
\newcommand{\I}{\mathrm{I}}
\newcommand{\II}{\mathrm{II}}
\newcommand{\Tr}{\operatorname{Tr}}          
\allowdisplaybreaks[1]
\begin{document}


\title{\textbf{
 Ergotropy of multipartite entangled states in the quantum atmosphere
 } } 
	
\author{Ruopu Sun and Xiaofen Huang}
\email{huangxf1206@163.com} 
\affiliation{School of Mathematics and Statistics, Hainan Normal University, Haikou 571158, China}


\begin{abstract}
	
We investigate the ergotropy of tripartite Greenberger-Horne-Zeilinger (GHZ) and W states in the quantum atmosphere of a Schwarzschild black hole, focusing on how the extractable work is redistributed under Hawking-induced thermalization. We find that the two classes of multipartite entanglement exhibit qualitatively different ergotropy structures. For GHZ states, the ergotropies associated with different accessible subsystems organize into two degenerate families, whereas for W states they obey a linear averaging relation with no counterpart in the GHZ case. Under symmetric gravitational thermal corrections, the two GHZ families remain separated by a constant energy gap that is independent of the radial position. We further show that the Hartle-Hawking constant $D_{HH}$ and the normalized radial distance $r/r_h$ play competing roles in the evolution of extractable work: increasing $D_{HH}$ suppresses its recovery toward the asymptotic Hawking limit, while increasing $r/r_h$ enhances the ergotropy accessible to the observers. These results show that the redistribution of extractable work in a black-hole environment is governed not only by gravitationally induced thermalization but also by the structure of the underlying multipartite entanglement, providing a thermodynamic signature that distinguishes GHZ- and W-type quantum correlations in curved spacetime.
    
\end{abstract}
	
\maketitle


\section{INTRODUCTION}
    
Multipartite entangled states provide a versatile platform for investigating quantum thermodynamics and the role of quantum correlations in thermodynamic processes. 
A central quantity in this context is the ergotropy, 
which quantifies the maximum work that can be extracted from a quantum state by arbitrary unitary transformations \cite{perciavalle2026quantumcoherence,lumbreras2025quantumstateagnostic,chen2025quantumworkmoving,biswas2025quantumthermodynamic,george2024workextractioncontrolled,williamson2024extractingwork}. 
As a fundamental thermodynamic resource, ergotropy has attracted considerable attention in recent years \cite{wang2025ergotropyquantumbatteries,cenedese2025boostingworkextractionquantum,thakuria2025coherenceleakstoragekurtosis,ahuja2025enhancingworkextractionquantumbatteries,zhang2025coherentquantumworkextraction,beder2026,chaki2025positivenonpositivemeasurementsenergy,andolina2019extractable,garciapintos2020fluctuations,delmonte2021characterization,liu2021entanglementcentralspin,shi2022entanglementcoherence}.

A natural extension of these considerations is to ask how ergotropy and other quantum resources are affected by relativistic motion and gravitational fields. Black-hole spacetimes provide a particularly interesting setting for this question. An observer hovering close to the event horizon experiences a thermal spectrum of particles, closely related to the Unruh effect for uniformly accelerated observers in flat spacetime. At the field-theoretic level, the corresponding Bogoliubov transformation mixes modes across the horizon, so that quantum information initially encoded in global field modes is redistributed between causally disconnected regions. Such a redistribution can substantially alter not only quantum correlations but also the amount of energy that remains extractable as useful work.

The relevant physics, however, is not necessarily confined to an infinitesimal neighborhood of the event horizon. Hawking quanta can be viewed as emerging from an extended near-horizon region—the so-called quantum atmosphere—whose characteristic size is comparable to the horizon radius. It is therefore natural to ask how quantum correlations and thermodynamic resources vary throughout this region, rather than only in the asymptotic Hawking limit. Studying their spatial redistribution provides a way to characterize the interplay between quantum information and thermodynamics in curved spacetime and may also offer insight into broader questions associated with Hawking radiation, including the information-loss and trans-Planckian problems.

This question becomes particularly relevant for a Schwarzschild black hole in the Hartle–Hawking vacuum  \cite{wu2024genuinely,moustos2025quantumthermalbtz,mikovic2024physicalstates,lewis2020classicalsimulations}, where the thermal environment experienced by a local observer depends explicitly on the radial position. 
In this setting, the asymptotic Hawking temperature is replaced by a local temperature determined by the radial coordinate and the Hartle–Hawking constant $D_{HH}$, providing a natural framework for probing quantum resources throughout the quantum atmosphere. 
Previous studies have mainly focused on how such horizon-induced thermal effects modify quantum correlations, showing, in particular, that multipartite entanglement can be degraded as the black hole is approached. 
The loss of entanglement, however, does not by itself determine how much useful energy remains stored in the quantum state. 
This distinction is captured by ergotropy, which quantifies the maximum work extractable from a state through unitary operations and thus provides a thermodynamic perspective complementary to correlation-based measures. 
It is also directly relevant to the operation of quantum batteries, heat engines, and energy-transfer protocols in relativistic settings. 
Despite this connection, little is known about how ergotropy is redistributed across the quantum atmosphere. In particular, the combined influence of radial position, the Hartle–Hawking correction to the local temperature, and multipartite entanglement on the extractable work has yet to be systematically explored.

Motivated by this gap, we investigate how multipartite entanglement influences the extractable work available to observers in the quantum atmosphere of a Schwarzschild black hole. 
Specifically, we consider GHZ-type \cite{zhou2021robustselftestingmultipartiteghzstate,menegassopires2021quantumcircuitsynthesis,majumdar2021quantumnetworkrecovery,musanna2020quantumsecretsharingusing,cao2019multipartyquantumprivacycomparison,cruz2019efficientquantumalgorithms} and W-type \cite{chatterjee2025deterministicdistributionwclassstates,liu2025identifying3qubitwstate,catalano2025experimentalpreparationwstates,farrell2025digitalquantumsimulationsscattering,kobayashi2026impossibilitywstates,li2024quantumsecretsharingenhanced,chen2022multipartiteconcurrencewclassstates} tripartite entangled states in Schwarzschild spacetime, incorporating the Hartle–Hawking correction to the local temperature \cite{hawking1974blackhole,denis2023entangled,almheiri2021entropy,wang2018hawking}. 
We derive the ergotropy associated with different partitions of the field modes and examine its dependence on the acceleration parameter, the normalized radial distance $r/r_h$, and the Hartle–Hawking constant $D_{HH}$. 
Our analysis reveals that the two classes of multipartite states respond to the gravitational thermal environment in qualitatively different ways. 
For GHZ-type states, the ergotropies organize into two degenerate families separated by a constant energy gap, whereas for W-type states they satisfy a linear averaging relation that is absent in the GHZ case. 
These distinct structures show that the redistribution of extractable work in the quantum atmosphere depends not only on the local thermal environment but also on how multipartite quantum information is encoded in the initial state. 
Our results thus establish a direct connection between multipartite entanglement structure, gravitationally induced thermalization, and quantum work extraction in curved spacetime.

The remainder of this paper is organized as follows. Section \ref{sec:Geometry_Schwarzschild} reviews the geometry of Schwarzschild spacetime and the vacuum structure of the Dirac field, where the Bogoliubov transformation relates the Kruskal and Schwarzschild modes. Section \ref{ERGOTROPY_ENTANGLED} is devoted to the ergotropy of multipartite entangled states in the black-hole quantum atmosphere. Section \ref{conclusion} summarizes our findings.


\section{Geometry of the Schwarzschild Spacetime}\label{sec:Geometry_Schwarzschild}

The Schwarzschild solution is the simplest vacuum black hole of general relativity, describing a static, uncharged, spherically symmetric spacetime. 
The Schwarzschild metric reads
\begin{equation}
\begin{split}
ds^2 &= -\left(1-\frac{2M}{r}\right)dt^2 + \left(1-\frac{2M}{r}\right)^{-1}dr^2 \\
&\quad + r^2d\theta^2 + r^2\sin^2\theta d\phi^2,
\end{split}
\label{eq:sch_metric}
\end{equation}
where $M$ denotes the black hole mass, $r$ is the radial coordinate, 
$t$ denotes the time coordinate, $\theta$ denotes the polar angle, 
and $\phi$ denotes the azimuthal angle. 
Also, we adopt natural units where $G = c = \hbar = k_B = 1$ for simplicity.

The Dirac equation in Schwarzschild spacetime reads
\begin{equation}
\left[\gamma^a e_a^\mu\left(\partial_\mu + \Gamma_\mu\right)\right]\psi = 0,
\end{equation}
where $\gamma^a$ denotes the Dirac matrices, $e^\mu_a$ represents the vierbein, 
and $\Gamma_\mu$ stands for the spin connection coefficients. 
Substituting the Schwarzschild vierbein into the Dirac equation, 
the Dirac equation for Schwarzschild spacetime can be expressed as
\begin{equation}
\begin{split}
&-\frac{\gamma_0}{\sqrt{1-\frac{2M}{r}}}\frac{\partial \psi}{\partial t}
+\gamma_1\sqrt{1-\frac{2M}{r}}\left[\frac{\partial}{\partial r}+\frac{1}{r}+\frac{M}{2r(r-2M)}\right]\psi \\
&+\frac{\gamma_2}{r}\left(\frac{\partial}{\partial \theta}+\frac{\cot\theta}{2}\right)\psi
+\frac{\gamma_3}{r\sin\theta}\frac{\partial \psi}{\partial \varphi}=0,
\end{split}
\label{eq:dirac_full}
\end{equation}
where $\gamma_k$ ($k=0,1,2,3$) are the Dirac matrices. 
This equation admits a set of positive-frequency outgoing solutions (Schwarzschild modes) localized inside and outside the event horizon, 
which are explicitly written as
\begin{equation}
\begin{cases}
\psi_k^{\I+} = \varsigma(r)e^{-i\omega u},\\
\psi_k^{\II+} = \varsigma(r)e^{i\omega u},
\end{cases}
\label{eq:fermi_modes}
\end{equation}
where $\psi_k^{\I+}$ and $\psi_k^{\II+}$ correspond to the positive-frequency solutions outside (Region $\I$) and inside (Region $\II$) the event horizon, $\omega$ denotes the monochromatic frequency of the Dirac field, 
$\varsigma(r)$ represents the four-component Dirac spinor, 
and the degenerate coordinate $u$ is defined as $u = t - r_*$, 
where $r_* = r + 2M\ln\left(\frac{r}{2M}-1\right)$.
    
Using the Damour--Ruffini method, the positive energy Kruskal modes bridge the two families of Schwarzschild modes in Eq.\eqref{eq:fermi_modes}:
\begin{equation}\label{eq:bogo_kruskal}
\begin{aligned}
&\Phi_{k,\I}^+= e^{-2\pi M\omega}\Psi_{-k,\II}^- + e^{2\pi M\omega}\Psi_{k,\I}^+, \\
&\Phi_{k,\II}^+ = e^{-2\pi M\omega}\Psi_{-k,\I}^- + e^{2\pi M\omega}\Psi_{k,\II}^+.
\end{aligned}
\end{equation}

Under the appropriate Kruskal modes, the Dirac field can be expanded as
\begin{equation}
\begin{split}
\psi = &\int dk\,\big[2\cosh(4\pi M\omega_i)\big]^{-\frac12}
\Big[\hat{c}_k^{\II}\Psi_{k,\II}^+ + \hat{d}_{-k}^{\II\dagger}\Psi_{-k,\II}^- \\
&+ \hat{c}_k^{\I}\Psi_{k,\I}^+ + \hat{d}_{-k}^{\I\dagger}\Psi_{-k,\I}^-\Big],
\end{split}
\label{eq:field_expansion}
\end{equation}
where the operators $\hat{c}_k$ and $\hat{d}_{-k}^\dagger$ serve as the creation and annihilation operators that operate on the Kruskal vacuum. 
    
From the Bogoliubov transform, the Kruskal vacuum state and excited state in Schwarzschild spacetime are expressed as
\begin{equation}
\begin{aligned}
|0\rangle_k &= \frac{1}{\sqrt{e^{-\frac{\omega}{T}} + 1}}|0\rangle_\mathrm{I}|0\rangle_\mathrm{II} + \frac{1}{\sqrt{e^{\frac{\omega}{T}} + 1}}|1\rangle_\mathrm{I}|1\rangle_\mathrm{II}, \\
|1\rangle_k &= |1\rangle_\mathrm{I}|0\rangle_\mathrm{II},
\end{aligned}
\label{eq:kruskal_vac}
\end{equation}
where $T = \frac{1}{8\pi M}$ denotes the Hawking temperature, 
$\omega$ is the particle frequency. 
To simplify the subsequent analysis, we define the substitution $\frac{1}{\sqrt{e^{-\frac{\omega}{T}} + 1}} = \cos\zeta$ and $\frac{1}{\sqrt{e^{\frac{\omega}{T}} + 1}} = \sin\zeta$, 
where $\zeta\in[0,\pi/4]$ is the thermal parameter.


\section{ERGOTROPY OF MULTIPARTITE ENTANGLED STATES 
IN THE BLACK HOLE QUANTUM ATMOSPHERE}\label{ERGOTROPY_ENTANGLED}	
     
\subsection{Definition of Ergotropy}\label{Definition_ Ergotropy}

Ergotropy, first introduced by Allahverdyan, Balian, 
and Nieuwenhuizen \cite{allahverdyan2004ergotropy}, 
is the maximal work extractable from a quantum state by arbitrary unitary operations and has become a central resource measure in quantum thermodynamics. 
It has since been investigated in diverse settings, 
ranging from the relation between entanglement and extractable work in Gaussian states \cite{lee2025quantumentanglement} to work extraction in quantum batteries \cite{ahuja2025enhancingworkextraction,beder2026workextraction}.

Consider a $d$-dimensional quantum system with density operator $\rho$ and Hamiltonian
\begin{equation}
H = \sum_{k=0}^{d-1} \epsilon_k |k\rangle\langle k|,
\label{eq:hamiltonian_spectral}
\end{equation}
where the energy levels are ordered as
$\epsilon_0\leqslant\epsilon_1\leqslant\cdots\leqslant\epsilon_{d-1}$ and $\{|k\rangle\}$ are the
corresponding orthonormal eigenstates. 
Let $\{\lambda_i\}$ be the eigenvalues of
$\rho$, ordered likewise, $\lambda_0\leqslant\lambda_1\leqslant\cdots\leqslant\lambda_{d-1}$.
Among all states unitarily equivalent to $\rho$, 
the unique state of minimum energy is the passive state
\begin{equation}
\rho^\downarrow = \sum_{i=0}^{d-1}\lambda_i|\epsilon_i\rangle\langle\epsilon_i|,
\label{eq:passive_state}
\end{equation}
which carries the same spectrum $\{\lambda_i\}$ as $\rho$. 
Ergotropy is then the energy gap between $\rho$ and its passive state,
\begin{equation}
\mathcal{E}(\rho) = \Tr[H\rho] - \Tr[H\rho^\downarrow],
\label{eq:ergotropy_def}
\end{equation}
with the passive-state energy
\begin{equation}
\Tr[H\rho^\downarrow]
= \sum_{i=0}^{d-1}\lambda_i\epsilon_{d-1-i}.
\label{eq:passive_energy}
\end{equation}

Ergotropy possesses several fundamental properties that characterize its role as a measure of extractable work. By definition, it is nonnegative, $\mathcal{E}(\rho)\geqslant 0$, and vanishes for passive states, including thermal (Gibbs) states, from which no work can be extracted through cyclic unitary operations. More importantly, ergotropy is determined not only by the populations of the energy levels but also by quantum coherence. Off-diagonal coherences in the energy eigenbasis can contribute to extractable work even in the absence of population inversion, thereby revealing an intrinsically quantum contribution to ergotropy. In multipartite systems, this connection is further enriched by correlations between the constituent subsystems, which can provide an additional source of extractable work. Consequently, the work accessible through global operations can exceed that obtainable through local operations, with the difference reflecting contributions associated with intersystem correlations, including entanglement. Ergotropy thus provides a thermodynamic framework for characterizing how multipartite quantum resources are converted into useful work and redistributed under gravitationally induced thermalization.

\subsection{ERGOTROPY OF THE GHZ TYPE
TRIPARTITE ENTANGLED STATES}\label{ERGOTROPY_GHZ}

To characterize ergotropy in a curved spacetime, 
we consider the generalized GHZ state. 
It provides a simple, tunable framework for analyzing how multipartite quantum resources evolve under the gravitational thermal effects of the black hole atmosphere.
In the initial inertial frame, the shared state reads
\begin{equation}
|\mathrm{GHZ}\rangle_{ABC}= \alpha|000\rangle + \sqrt{1-\alpha^{2}}|111\rangle,
\label{eq:ghz_initial}
\end{equation}
where $\alpha\in[0,1]$ controls the relative weight of the two GHZ branches~\cite{greenberger1990bell}.

For the tripartite system composed of subsystems $A$, $B$, and $C$, 
we consider an anisotropic three-body interaction Hamiltonian of the form
\begin{equation}
H =J_{\perp}
\left(
\sigma_x^{A}\sigma_x^{B}\sigma_x^{C}
+
\sigma_y^{A}\sigma_y^{B}\sigma_y^{C}
\right)
+
J_z \sigma_z^{A}\sigma_z^{B}\sigma_z^{C},
\label{eq:xyz_general}
\end{equation}
where \(\sigma_{j}^{\nu}\) (\(j=x,y,z\) and
\(\nu=A,B,C\)) denotes the Pauli operator acting on subsystem \(\nu\).
The coupling constants $J_{\perp}$ and $J_z$ characterize the strengths 
of the transverse and longitudinal three-body interactions, respectively.
The transverse terms generally induce mixing between different 
computational-basis states, whereas the longitudinal term is diagonal 
in the computational basis.

To focus on the change in extractable work associated with the population 
redistribution induced by the gravitational environment, we restrict the 
analysis to the purely longitudinal interaction. 
This choice avoids additional basis-state mixing generated by the transverse terms and allows the gravitational contribution to the ergotropy to be identified more directly. 
Accordingly, we set $J_{\perp}=0$ and take $J_z$ as the characteristic energy scale. 
Expressing all energies in units of $J_z$, 
the Hamiltonian used throughout the following analysis reduces to
\begin{equation}
H =\sigma_z^{A}\sigma_z^{B}\sigma_z^{C}.
\label{eq:h_ising}
\end{equation}

The Hamiltonian in Eq.~(\ref{eq:h_ising}) is diagonal in the computational basis and possesses two fourfold-degenerate eigenvalues. 
For the evaluation of ergotropy, we arrange the energy eigenvalues in nondecreasing order,
\begin{equation}
\epsilon_0=\epsilon_1=\epsilon_2=\epsilon_3=-1,
\qquad
\epsilon_4=\epsilon_5=\epsilon_6=\epsilon_7=+1.
\label{eq:energy_spectrum}
\end{equation}
Here, the indices $i=0,\ldots,7$ specify the ordering of the energy 
eigenvalues and will be used consistently in the subsequent ergotropy 
analysis. 
All ergotropies are correspondingly expressed in units of $J_z$.
This spectral structure provides a convenient and physically transparent 
framework for examining how gravitationally induced changes in the quantum 
state modify the amount of extractable work.

At first, Alice, Bob, and Charlie share the tripartite GHZ state
in the asymptotically flat region. 
Thereafter, Alice remains inertial in this region, 
while Bob and Charlie free fall into the Schwarzschild black hole 
and experience the Hawking thermal effect in its quantum atmosphere. 
Consequently, the initial state $|\mathrm{GHZ}\rangle_{ABC}$ evolves
into the five-partite state $|\mathrm{GHZ}\rangle_{AB_{\mathrm I}B_{\mathrm{II}}
C_{\mathrm I}C_{\mathrm{II}}}$, which follows from the Bogoliubov
transformation in Eq.~\eqref{eq:kruskal_vac} and reads explicitly

\begin{equation}
\begin{aligned}
|\mathrm{GHZ}&\rangle_{AB_{\mathrm{I}} B_{\mathrm{II}} C_{\mathrm{I}} C_{\mathrm{II}}}
= \alpha \sin\zeta_{b}\sin\zeta_{c}\left|01111\right\rangle \\
&+\alpha \cos\zeta_{b} \cos\zeta_{c}\left|00000\right\rangle 
+\alpha \cos\zeta_{b} \sin\zeta_{c}\left|00011\right\rangle \\
&+\alpha \sin\zeta_{b} \cos\zeta_{c}\left|01100\right\rangle 
+\sqrt{1-\alpha^{2}}\left|11010\right\rangle,
\end{aligned}
\label{eq:full_wavefunction}
\end{equation}
where $\zeta_b$ and $\zeta_c$ characterize the gravitational thermal
effects sensed by Bob and Charlie, respectively.
Owing to the causal separation
between Region $\I$ (outside the horizon) and Region $\II$ (inside the
horizon), 
only the modes $B_{\mathrm I}$ and $C_{\mathrm I}$ are
physically accessible to exterior observers, 
whereas the interior modes $B_{\mathrm{II}}$ and $C_{\mathrm{II}}$ remain inaccessible. 
Tracing out all unobservable degrees of freedom in Region $\II$ yields the reduced state of the accessible subsystem,
\begin{equation}
\rho^{\scriptscriptstyle\mathrm{GHZ}}_{AB_{\mathrm I} C_{\mathrm I}}=
\begin{pmatrix}
a_{1} & 0 & 0 & 0 & 0 & 0 & 0 & z_{1} \\
0 & a_{2} & 0 & 0 & 0 & 0 & 0 & 0 \\
0 & 0 & a_{3} & 0 & 0 & 0 & 0 & 0 \\
0 & 0 & 0 & a_{4} & 0 & 0 & 0 & 0 \\
0 & 0 & 0 & 0 & 0 & 0 & 0 & 0 \\
0 & 0 & 0 & 0 & 0 & 0 & 0 & 0 \\
0 & 0 & 0 & 0 & 0 & 0 & 0 & 0 \\
z_{1} & 0 & 0 & 0 & 0 & 0 & 0 & b_{1}
\end{pmatrix},
\end{equation}
where the matrix elements are given by

\begin{equation}
\begin{aligned}
b_{1}&=1-\alpha^{2},\quad b_{2,3,4}=0, \\
z_{1}&=\alpha\sqrt{1-\alpha^{2}}\cos\zeta_{b}\cos\zeta_{c}, \quad z_{2,3,4}=0,\\
a_{1}&=\alpha^{2}\cos^{2}\zeta_{b}\cos^{2}\zeta_{c}, \quad 
a_{2}=\alpha^{2}\cos^{2}\zeta_{b}\sin^{2}\zeta_{c}, \\
a_{3}&=\alpha^{2}\sin^{2}\zeta_{b}\cos^{2}\zeta_{c}, \quad 
a_{4}=\alpha^{2}\sin^{2}\zeta_{b}\sin^{2}\zeta_{c}, \\
\end{aligned}
\label{eq:rho_ABICI_elements}
\end{equation}

 And the eigenvalues of $\rho^{\scriptscriptstyle\mathrm{GHZ}}_{AB_{\mathrm I} C_{\mathrm I}}$  are as follows:
\begin{equation}
\begin{aligned}
\lambda_{0}&=\lambda_{1}=\lambda_{2}=\lambda_{3}=0,\\
\lambda_{4}&=\alpha^2\cos^2 \zeta_b\sin^2 \zeta_c,\\
\lambda_{5}&=\alpha^2\sin^2 \zeta_b\cos^2 \zeta_c,\\
\lambda_{6}&=\alpha^2\sin^2 \zeta_b\sin^2 \zeta_c,\\
\lambda_{7}&=1-\alpha^2\left(1-\cos^2\zeta_b\cos^2\zeta_c\right).
\end{aligned}
\label{eq:rho_ABICI_eigenvalues}
\end{equation}

Similarly, one can obtain the reduced density operators 
$\rho^{\scriptscriptstyle\mathrm{GHZ}}_{AB_{\mathrm I}C_{\mathrm{II}}}$,
$\rho^{\scriptscriptstyle\mathrm{GHZ}}_{AB_{\mathrm{II}}C_{\mathrm I}}$, $\rho^{\scriptscriptstyle\mathrm{GHZ}}_{AB_{\mathrm{II}}C_{\mathrm{II}}}$ and their corresponding eigenvalues, which are provided in Appendix \ref{app_GHZ}.

\begin{figure}
\centering
\includegraphics[width=0.72\linewidth]{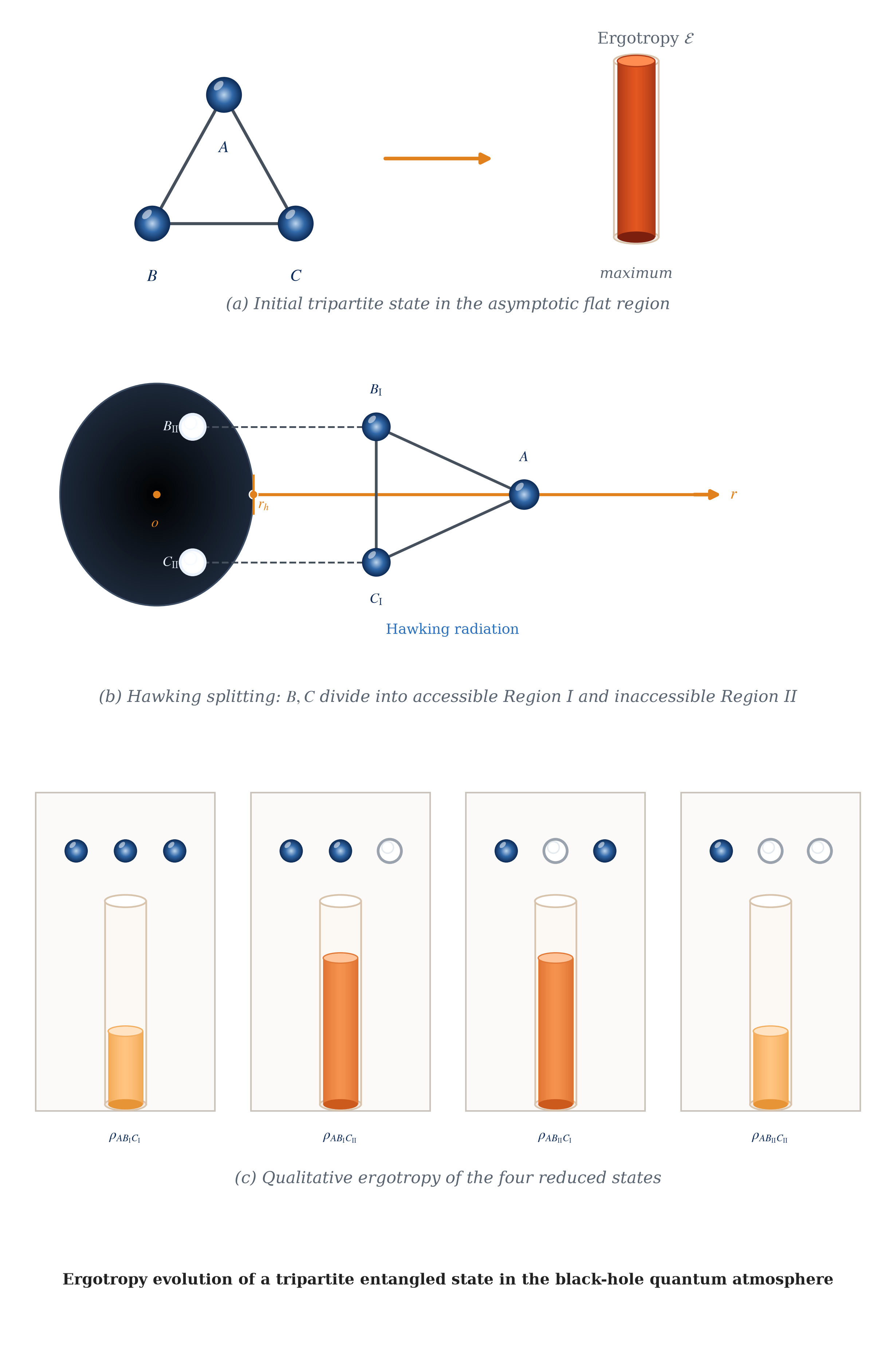}
\captionsetup{justification=raggedright,singlelinecheck=false}
\caption{Schematic of ergotropy evolution for a tripartite entangled state in the black-hole quantum atmosphere. 
(a)~In the asymptotic flat region the initial state possesses maximal ergotropy.
(b)~Via the Hawking (Bogoliubov) transformation, the accelerated modes $B$, $C$ split into an accessible Region~$I$ outside the horizon and a causally inaccessible Region~$II$ inside it, while Alice stays in the flat region. 
(c)~The four $I/II$ combinations yield four reduced states whose ergotropy forms two degenerate pairs.}
\label{fig:ergotropyschematicv3}
\end{figure}

The ergotropy in Eq.~\eqref{eq:ergotropy_def} is evaluated through the
passive state $\rho^\downarrow$. For the working Hamiltonian
$H=\sigma_z^{A}\sigma_z^{B}\sigma_z^{C}$, the spectrum contains only two
distinct values, $-1$ and $+1$, each fourfold degenerate. Throughout the
parameter range $\alpha\in[0,1]$ and
$0<\zeta_b=\zeta_c=\zeta\leqslant\pi/4$, four of the eight eigenvalues of
each reduced state are identically zero, while the remaining four are
strictly positive. Consequently, the four positive eigenvalues are assigned
to the lower level $\epsilon=-1$ and the four zero eigenvalues to the upper
level $\epsilon=+1$.
With the spectra of
$\rho^{\scriptscriptstyle\mathrm{GHZ}}_{AB_{\mathrm I}C_{\mathrm I}}$,
$\rho^{\scriptscriptstyle\mathrm{GHZ}}_{AB_{\mathrm I}C_{\mathrm{II}}}$,
$\rho^{\scriptscriptstyle\mathrm{GHZ}}_{AB_{\mathrm{II}}C_{\mathrm I}}$,
and
$\rho^{\scriptscriptstyle\mathrm{GHZ}}_{AB_{\mathrm{II}}C_{\mathrm{II}}}$
at hand, their ergotropies
follow directly from Eq.~\eqref{eq:ergotropy_def}; 
the results are (illustrated schematically in FIG.~\ref{fig:ergotropyschematicv3}):

\begin{align}
\mathcal{E}\!\left(\rho^{\scriptscriptstyle\mathrm{GHZ}}_{AB_\mathrm{I}C_\mathrm{I}},H\right)
=\mathcal{E}\!\left(\rho^{\scriptscriptstyle\mathrm{GHZ}}_{AB_\mathrm{II}C_\mathrm{II}},H\right)
&= \alpha^2\left(\cos^2 2\zeta+1\right),
\label{eq:ergotropy_ABICI_ABIICII}\\
\mathcal{E}\!\left(\rho^{\scriptscriptstyle\mathrm{GHZ}}_{AB_\mathrm{I}C_\mathrm{II}},H\right)
=\mathcal{E}\!\left(\rho^{\scriptscriptstyle\mathrm{GHZ}}_{AB_\mathrm{II}C_\mathrm{I}},H\right)
&= \alpha^2\left(\cos^2 2\zeta-1\right)+2.
\label{eq:ergotropy_ABICII_ABIICI}
\end{align}

Substituting the above analytical expressions, we derive several intrinsic relations among the four ergotropies. 
The two groups of ergotropies have a fixed unit difference independent of parameters $\zeta_b$ and $\zeta_c$,
\begin{equation}
\mathcal{E}(\rho^{\scriptscriptstyle\mathrm{GHZ}}_{AB_\mathrm{I}C_\mathrm{II}},H)-\mathcal{E}(\rho^{\scriptscriptstyle\mathrm{GHZ}}_{AB_\mathrm{I}C_\mathrm{I}},H)=2-2\alpha^2.
\end{equation}

\begin{figure*}[t]
\centering
\subfloat[]{%
\includegraphics[width=0.22\textwidth]{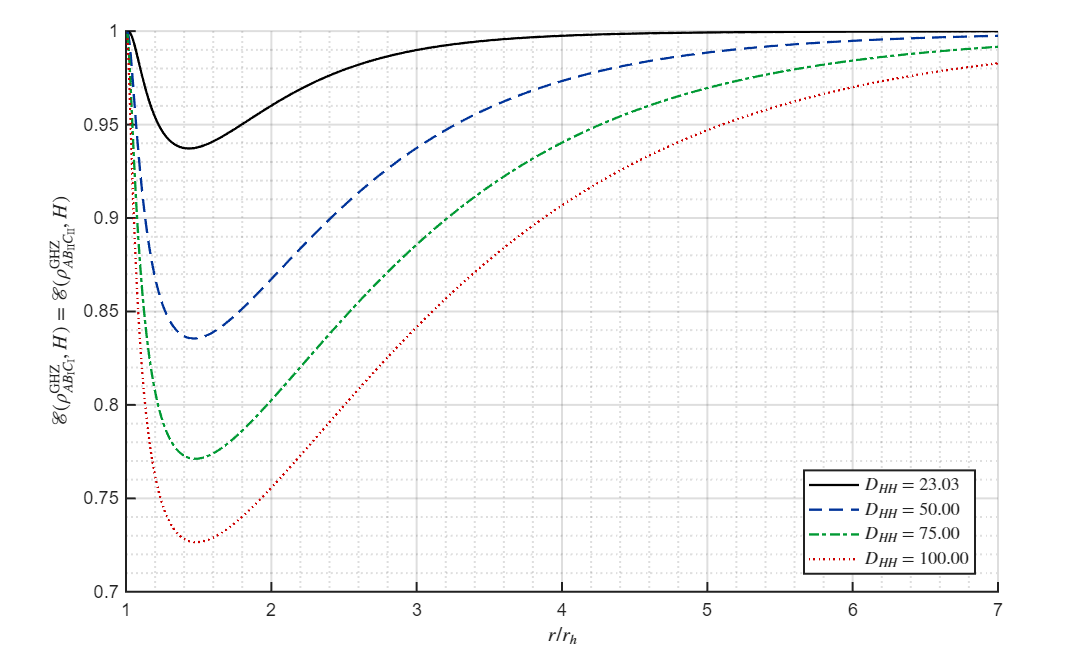}%
\label{fig:ghz_e_ab1c1_dhh}%
}
\hfill
\subfloat[]{%
\includegraphics[width=0.22\textwidth]{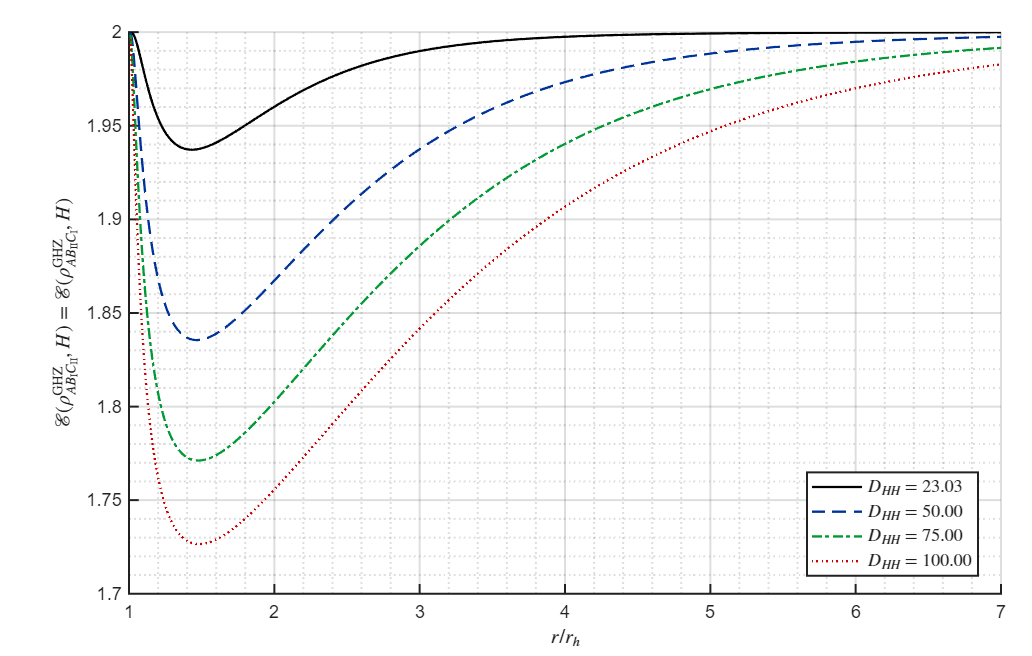}%
\label{fig:ghz_e_ab1c2_dhh}%
}
\hfill
\subfloat[]{%
\includegraphics[width=0.22\textwidth]{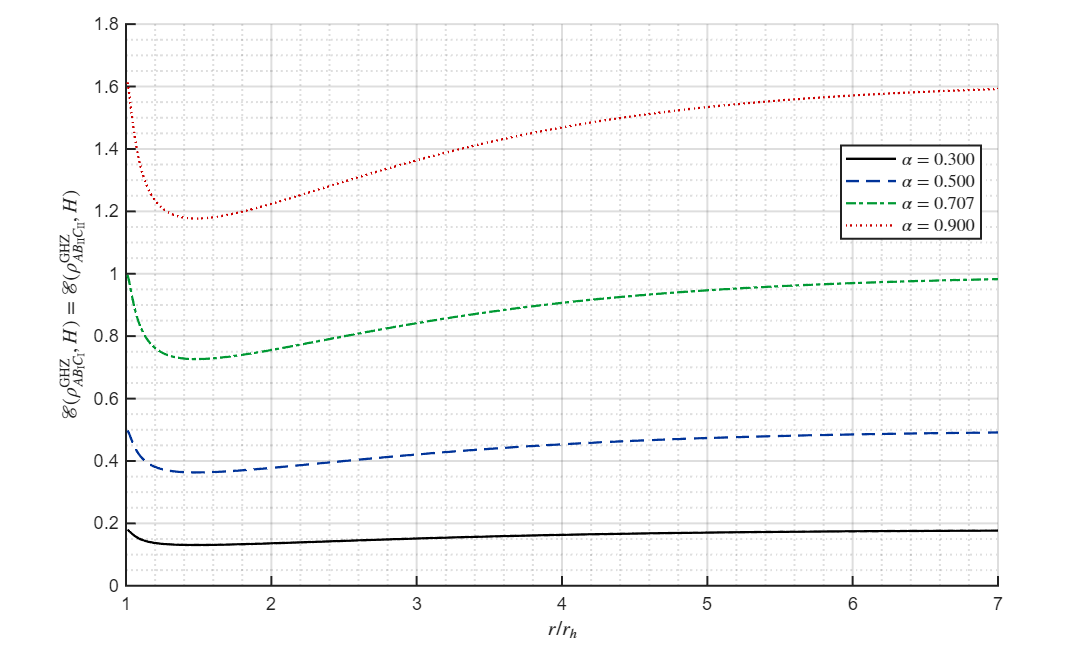}%
\label{fig:ghz_e_ab1c1_alpha}%
}
\hfill
\subfloat[]{%
\includegraphics[width=0.22\textwidth]{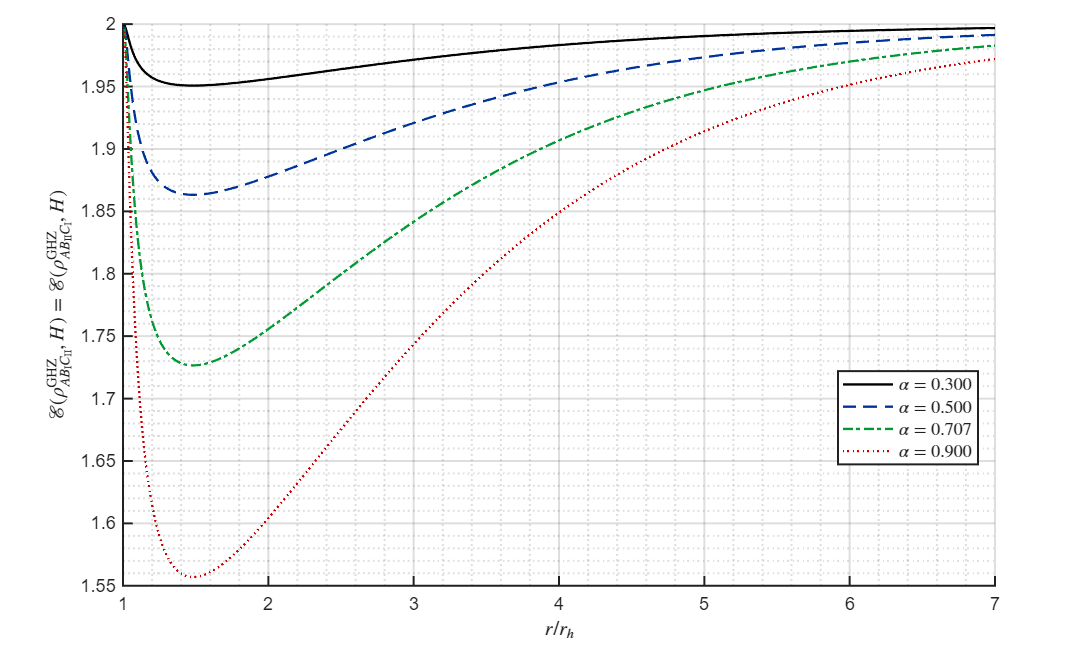}%
\label{fig:ghz_e_ab1c2_alpha}%
}
\caption{The ergotropy of the $\rho^{\scriptscriptstyle\mathrm{GHZ}}_{AB_{\mathrm I}C_{\mathrm I}}$,
$\rho^{\scriptscriptstyle\mathrm{GHZ}}_{AB_{\mathrm I}C_{\mathrm{II}}}$,
$\rho^{\scriptscriptstyle\mathrm{GHZ}}_{AB_{\mathrm{II}}C_{\mathrm I}}$,
and
$\rho^{\scriptscriptstyle\mathrm{GHZ}}_{AB_{\mathrm{II}}C_{\mathrm{II}}}$ as a function of the normalized radial distance $r/r_h$ in the black hole quantum
atmosphere.
\textbf{(a),(b)}~Fixed at $\alpha=1/\sqrt{2}$ for
different Hartle--Hawking constants $D_{HH}$;
\textbf{(c),(d)}~fixed at $D_{HH}=100$ for different $\alpha$.}
\label{fig:ghz_ergotropy_atmosphere}
\end{figure*}

To incorporate the effect of the black-hole atmosphere, 
we replace the Hawking temperature $T$ entering the Bogoliubov coefficients with the local temperature
$T_{HH}$ in the Hartle--Hawking vacuum, which is given by
\begin{equation}
\begin{aligned}
T_{HH} &= T_H \sqrt{1-\frac{r_h}{r}} \\
&\quad \cdot \sqrt{1+2\frac{r_h}{r}+\left(\frac{r_h}{r}\right)^2
\left(9+4D_{HH}+36\ln\left(\frac{r_h}{r}\right)\right)},
\end{aligned}
\label{eq:THH}
\end{equation}
where $T_H = 1/(4\pi r_h)$~\cite{eune2019,kaczmarek2024,kaczmarek240508167}, 
and $D_{HH}$ is the stress-tensor constant associated with the Hartle--Hawking
vacuum. 
Since the Hartle--Hawking boundary conditions alone cannot fix
$D_{HH}$, additional conditions are required. 
To keep the temperature real everywhere outside the horizon and to rule out an unphysical inverse dependence on distance, 
we restrict ourselves to $D_{HH}\geqslant D_c\simeq 23.03$; 
under this condition $T_{HH}$ attains a maximum at $r_c\simeq 1.43\,r_h$ and decreases monotonically as $r$ increases beyond $r_c$. 
We adopt $D_{HH}\geqslant D_c$ throughout. 
Moreover, the local temperature vanishes at the horizon
($r=r_h$) and approaches the Hawking temperature as $r\to\infty$
~\cite{eune2019test,kaczmarek2024signatures,kaczmarek2026nonlocal,zhang2025entanglement}. 
Through the Eqs. \eqref{eq:ergotropy_ABICI_ABIICII}, \eqref{eq:ergotropy_ABICII_ABIICI} and \eqref{eq:THH}, the corresponding ergotropies are illustrated in  FIG.~\ref{fig:ghz_ergotropy_atmosphere}.

The $\mathcal{E}\!\left(\rho^{\scriptscriptstyle\mathrm{GHZ}}_{AB_{\mathrm I}C_{\mathrm I}},H\right)=\mathcal{E}\!\left(\rho^{\scriptscriptstyle\mathrm{GHZ}}_{AB_{\mathrm{II}}C_{\mathrm{II}}},H\right)$
and
$\mathcal{E}\!\left(\rho^{\scriptscriptstyle\mathrm{GHZ}}_{AB_{\mathrm I}C_{\mathrm{II}}},H\right)=\mathcal{E}\!\left(\rho^{\scriptscriptstyle\mathrm{GHZ}}_{AB_{\mathrm{II}}C_{\mathrm I}},H\right)$
exhibit a uniform nonmonotonic radial profile: 
each starts from its maximal value at the horizon, dips to a minimum in the near-horizon region, and then recovers toward the asymptotic Hawking limit at large
$r/r_h$. 
The minimum coincides with the maximum of the local temperature
$T_{HH}$, confirming that the radial redistribution of extractable work is
directly driven by the atmospheric thermal peak. 
For fixed $\alpha=1/\sqrt{2}$ [FIG.~\ref{fig:ghz_ergotropy_atmosphere}(a),(b)],
increasing $D_{HH}$ deepens the near-horizon minimum of both
$\mathcal{E}(\rho^{\scriptscriptstyle\mathrm{GHZ}}_{AB_{\mathrm I}C_{\mathrm I}})$
and
$\mathcal{E}(\rho^{\scriptscriptstyle\mathrm{GHZ}}_{AB_{\mathrm I}C_{\mathrm{II}}})$,
while simultaneously shifting the extremal position slightly outward and
slowing the large-$r$ recovery. Thus a larger Hartle--Hawking constant
strengthens the near-horizon thermal suppression and delays the return to
the flat-space value. For fixed $D_{HH}=100$
[FIG.~\ref{fig:ghz_ergotropy_atmosphere}(c),(d)],
varying $\alpha$ reshapes both curves systematically. 
As $\alpha$ increases, the overall level of
$\mathcal{E}(\rho^{\scriptscriptstyle\mathrm{GHZ}}_{AB_{\mathrm I}C_{\mathrm I}})$ rises and its near-horizon minimum deepens, 
evolving from an almost flat profile at small $\alpha$ to a pronounced dip at large $\alpha$. 
For $\mathcal{E}(\rho^{\scriptscriptstyle\mathrm{GHZ}}_{AB_{\mathrm I}C_{\mathrm{II}}})$,
the large-$r$ plateau remains unchanged for all $\alpha$, 
but its near-horizon minimum likewise deepens as $\alpha$ grows.

\subsection{ERGOTROPY OF THE W TYPE TRIPARTITE ENTANGLED STATES}
To further investigate the properties of ergotropy, we additionally choose the W state as the initial state and compare the evolution of its extractable work in the black hole with that of the GHZ state.

we assume that Alice, Bob, and Charlie all reside in the asymptotically flat region and share a tripartite W-type entangled state~\cite{dur2000three,li2025acceleration}, which reads
\begin{align}
|W\rangle_{ABC}
=\frac{1}{\sqrt{3}}
\Big(\big|001\big\rangle
+\big|010\big\rangle
+\big|100\big>\Big).
\end{align}
Thereafter, Alice remains stationary within this flat region, while Bob and Charlie hover near the event horizon with constant proper acceleration and are subjected to the thermal effects of Hawking radiation in the black hole quantum atmosphere outside $r_h$. Accordingly, the original $|W\rangle_{ABC}$ will be transformed into a five-partite quantum state $|W\rangle_{A B_\mathrm{I} B_\mathrm{II} C_\mathrm{I} C_\mathrm{II}}$, whose analytical expression is given as
\begin{equation}
\begin{split}
&|W\rangle_{A B_\mathrm{I} B_\mathrm{II} C_\mathrm{I} C_\mathrm{II}}= \\
&\frac{1}{\sqrt{3}} \Big(
\xi |00010\rangle + \eta |01110\rangle
 + \xi |01000\rangle+ \eta |01010\rangle\\
&+\xi^2 |10000\rangle
 + \xi\eta |10011\rangle
+ \xi\eta |11100\rangle + \eta^2 |11111\rangle
\Big).
\end{split}
\label{eq:w-fullstate}
\end{equation}
where~\( \xi = \left(e^{-\omega/T}+1\right)^{-1/2}\), \(\eta = \left(e^{\omega/T}+1\right)^{-1/2}\), $\xi^2 + \eta^2 = 1$, 
which is used in the subsequent simplification. 
Given the causal disconnection between Region $I$ and Region $II$, and the fact that exterior observers cannot access the field modes confined within the event horizon, modes $B_\mathrm{I}$ and $C_\mathrm{I}$ located outside the horizon are referred to as physically accessible, whereas interior modes $B_\mathrm{II}$ and $C_\mathrm{II}$ situated inside the event horizon are designated as inaccessible. 
Through tracing over all degrees of freedom belonging to Region $II$, we can extract the state of the physically accessible subsystem consisting of Alice, $B_\mathrm{I}$ and $C_\mathrm{I}$, thereby leading to the reduced density matrix $\rho_{AB_\mathrm{I}C_\mathrm{I}}$

\begin{equation}
\begin{aligned}
\rho^{\scriptscriptstyle\mathrm{W}}_{A B_\mathrm{I} C_\mathrm{I}}
=\frac{1}{3}
\begin{pmatrix}
0 & 0 & 0 & 0 & 0 & 0 & 0 & 0 \\
0 & \xi^2 & \xi^2 & 0 & \xi^3 & 0 & 0 & 0 \\
0 & \xi^2 & \xi^2 & 0 & \xi^3 & 0 & 0 & 0 \\
0 & 0 & 0 & 2\eta^2 & 0 & \xi\eta^2 & \xi\eta^2 & 0 \\
0 & \xi^3 & \xi^3 & 0 & \xi^4 & 0 & 0 & 0 \\
0 & 0 & 0 & \xi\eta^2 & 0 & \xi^2\eta^2 & 0 & 0 \\
0 & 0 & 0 & \xi\eta^2 & 0 & 0 & \xi^2\eta^2 & 0 \\
0 & 0 & 0 & 0 & 0 & 0 & 0 & \eta^4
\end{pmatrix}.
\end{aligned}
\end{equation}

The $\rho^{\scriptscriptstyle\mathrm{W}}_{A B_\mathrm{I} C_\mathrm{I}}$, $\rho^{\scriptscriptstyle\mathrm{W}}_{A B_\mathrm{I} C_\mathrm{II}}$, $\rho^{\scriptscriptstyle\mathrm{W}}_{A B_\mathrm{II} C_\mathrm{I}}$, $\rho^{\scriptscriptstyle\mathrm{W}}_{A B_\mathrm{II} C_\mathrm{II}}$ 
and their corresponding eigenvalues are provided in Appendix B.
	
The four reduced states
$\rho^{\scriptscriptstyle\mathrm{W}}_{AB_{\mathrm I}C_{\mathrm I}}$,
$\rho^{\scriptscriptstyle\mathrm{W}}_{AB_{\mathrm I}C_{\mathrm{II}}}$,
$\rho^{\scriptscriptstyle\mathrm{W}}_{AB_{\mathrm{II}}C_{\mathrm I}}$,
and
$\rho^{\scriptscriptstyle\mathrm{W}}_{AB_{\mathrm{II}}C_{\mathrm{II}}}$ share a common spectral structure: 
exactly four of their eight eigenvalues vanish identically, 
while the remaining four are strictly positive. 
For comparison, we use the same Hamiltonian as in the GHZ case. Under the two four-fold-degenerate levels of $H =\sigma_z^{A}\sigma_z^{B}\sigma_z^{C}$, 
the passive state therefore assigns the four positive eigenvalues to the lower level $\epsilon=-1$ and the four zeros
to the upper level $\epsilon=+1$, 
so that the passive-state energy is determined solely by the four positive eigenvalues. With this ordering,
the ergotropy of each reduced W state follows directly from
Eq.~\eqref{eq:ergotropy_def}.
\setlength{\jot}{2pt}
\begin{equation}
\begin{aligned}
&\mathcal{E}(\rho^{\scriptscriptstyle\mathrm{W}}_{AB_\mathrm{I}C_\mathrm{I}},H) =\frac{1}{3}\bigl(8\eta^2 - 4\eta^4\bigr), \\
&\mathcal{E}(\rho^{\scriptscriptstyle\mathrm{W}}_{AB_\mathrm{II}C_\mathrm{II}},H) = \frac{1}{3}\bigl(4 - 4\eta^4\bigr), \\
&\mathcal{E}(\rho^{\scriptscriptstyle\mathrm{W}}_{AB_\mathrm{I}C_\mathrm{II}},H) =\mathcal{E}(\rho^{\scriptscriptstyle\mathrm{W}}_{AB_\mathrm{II}C_\mathrm{I}},H)  =\frac{1}{3}\bigl(-4\eta^4 + 4\eta^2 + 2\bigr). \\
\end{aligned}
\label{eq:all_erg}
\end{equation}
\setlength{\jot}{6pt} 
	
It can be found that the ergotropies of the three reduced states satisfy a strict linear relation. Specifically, the ergotropy of $\rho^{\scriptscriptstyle\mathrm{W}}_{AB_\mathrm{I}C_\mathrm{II}}$($\rho^{\scriptscriptstyle\mathrm{W}}_{AB_\mathrm{II}C_\mathrm{I}}$) equals the arithmetic mean of the ergotropies of $\rho^{\scriptscriptstyle\mathrm{W}}_{AB_\mathrm{I}C_\mathrm{I}}$ and $\rho^{\scriptscriptstyle\mathrm{W}}_{AB_\mathrm{II}C_\mathrm{II}}$, i.e.,
\begin{equation}
\mathcal{E}(\rho^{\scriptscriptstyle\mathrm{W}}_{AB_\mathrm{I}C_\mathrm{II}},H)=\frac{1}{2}\Big[\mathcal{E}(\rho^{\scriptscriptstyle\mathrm{W}}_{AB_\mathrm{I}C_\mathrm{I}},H)
+\mathcal{E}(\rho^{\scriptscriptstyle\mathrm{W}}_{AB_\mathrm{II}C_\mathrm{II}},H)\Big].
\end{equation}

\begin{figure*}[htbp]
\centering
\subfloat[]{%
\includegraphics[width=0.32\linewidth]{"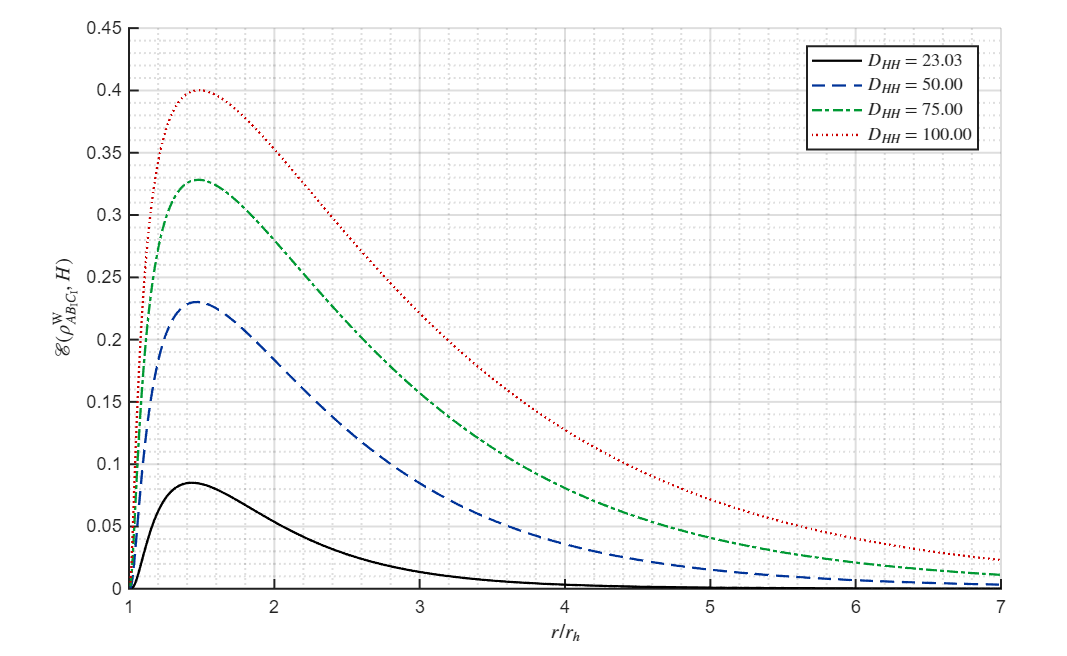"}%
\label{fig:6a}%
}
\hfill
\subfloat[]{%
\includegraphics[width=0.32\linewidth]{"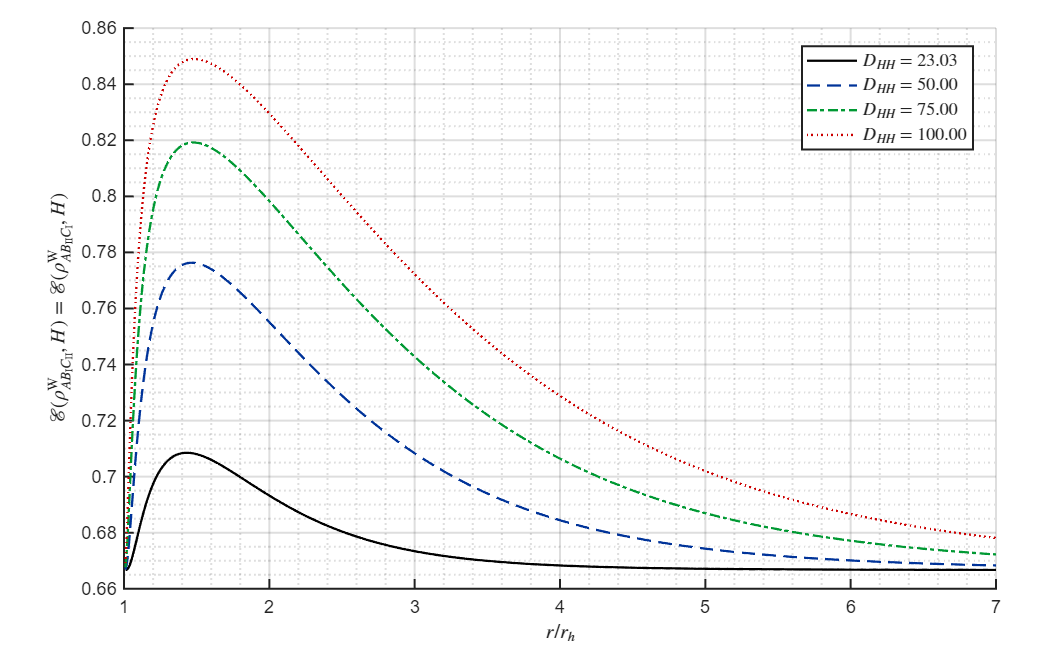"}%
\label{fig:6b}%
}
\hfill
\subfloat[]{%
\includegraphics[width=0.32\linewidth]{"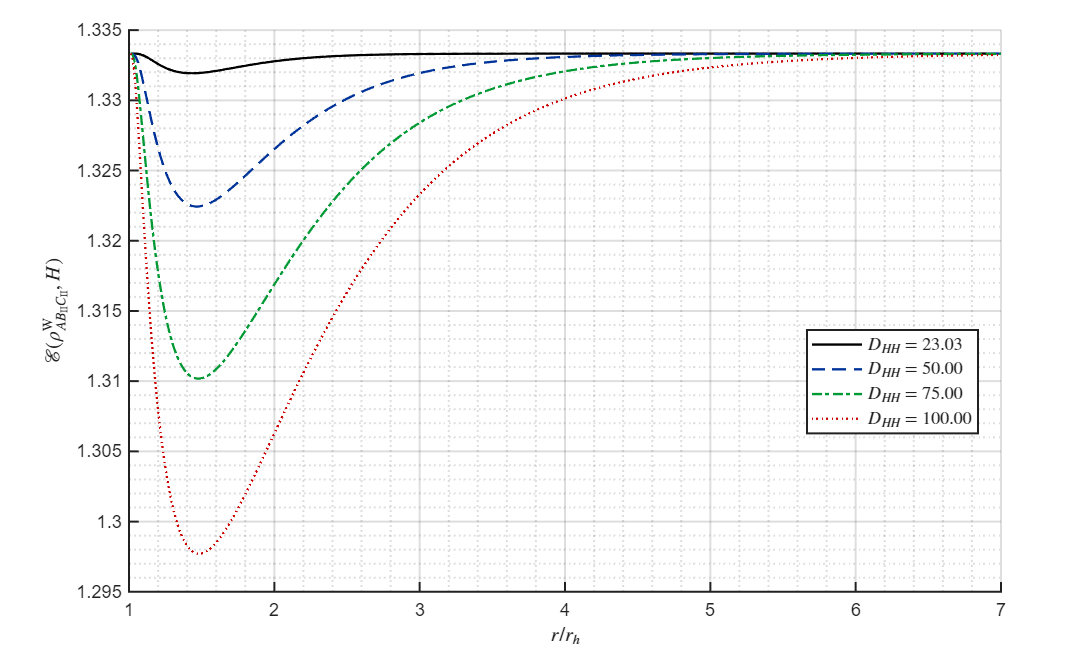"}%
\label{fig:6c}%
}
\caption{The ergotropy of three bipartite subsystems of the W state as a function of normalized radial distance $r/r_h$ under different Hartle-Hawking constants $D_{HH}$. 
Graph (a): ergotropy $\mathcal{E}(\rho^{\scriptscriptstyle\mathrm{W}}_{AB_\mathrm{I}C_\mathrm{I}},H)$;
Graph (b): ergotropy $\mathcal{E}(\rho^{\scriptscriptstyle\mathrm{W}}_{AB_\mathrm{I}C_\mathrm{II}},H)=\mathcal{E}(\rho^{\scriptscriptstyle\mathrm{W}}_{AB_\mathrm{II}C_\mathrm{I}},H)$; 
Graph (c): ergotropy $\mathcal{E}(\rho^{\scriptscriptstyle\mathrm{W}}_{AB_\mathrm{II}C_\mathrm{II}},H)$.}
\label{fig:5}
\end{figure*}

Meanwhile, we use the analogous manner, 
through the Eqs. \eqref{eq:all_erg} and \eqref{eq:THH}, 
the corresponding ergotropies are illustrated in  FIG.~\ref{fig:5}, the ergotropy of the
accessible and inaccessible reductions of the tripartite W~state exhibits
pronounced local extrema as a function of $r/r_h$. For the fully
accessible sector
$\mathcal{E}\!\left(\rho^{\scriptscriptstyle\mathrm{W}}_{AB_{\mathrm I}C_{\mathrm I}},H\right)$,
the ergotropy starts from zero at the horizon, rises into a single
localized peak in the near-horizon region, and decays back to zero at
large $r/r_h$. The
mixed sector
$\mathcal{E}\!\left(\rho^{\scriptscriptstyle\mathrm{W}}_{AB_{\mathrm I}C_{\mathrm{II}}},H\right)
=\mathcal{E}\!\left(\rho^{\scriptscriptstyle\mathrm{W}}_{AB_{\mathrm{II}}C_{\mathrm I}},H\right)$
shows the same atmosphere-induced enhancement: it departs from a common
value at the horizon, rises into a coincident peak in the near-horizon
region, and settles back to that same value at infinity. In contrast, the
fully inaccessible ergotropy
$\mathcal{E}\!\left(\rho^{\scriptscriptstyle\mathrm{W}}_{AB_{\mathrm{II}}C_{\mathrm{II}}},H\right)$
follows the opposite trend, taking identical values at the horizon and at
infinity but dipping into a minimum at the same near-horizon scale. This
mirror-like behavior shows that the local Hawking effect redistributes
extractable work between accessible and inaccessible modes with opposite
signs.
.
	
To better illustrate the radial dependence of the extractable work for
the two inequivalent multipartite entangled states, 
we summarize their extremal positions and values in TABLE~\ref{tab:extrema_compare}.
As $D_{HH}$ increases from $23.03$ to $100$, 
the extrema shift outward from $r_{\mathrm{ext}}/r_h\simeq1.43$ to approximately $1.48$, 
remaining within the narrow near-horizon interval
$1.43\lesssim r/r_h<1.5$.
The extremal positions of the GHZ and W states nearly coincide,
indicating that the ergotropy of both states is sensitive to the local
Hawking effect in the same radial region, 
tracking the maximum of the Hartle--Hawking temperature $T_{HH}$.
However, the two states respond differently to increasing $D_{HH}$:
the GHZ ergotropy decreases monotonically for both families, maintaining
the parameter-independent gap $2-2\alpha^{2}$, whereas the W state
exhibits a mixed response, with the maxima of
$\rho^{\scriptscriptstyle\mathrm{W}}_{AB_{\mathrm I}C_{\mathrm I}}$ and
$\rho^{\scriptscriptstyle\mathrm{W}}_{AB_{\mathrm I}C_{\mathrm{II}}}$ growing while the minimum of
$\rho^{\scriptscriptstyle\mathrm{W}}_{AB_{\mathrm{II}}C_{\mathrm{II}}}$ shrinks.
Thus, despite sharing the same near-horizon extremal region, 
the GHZ and W states provide complementary signatures of the local Hawking effect,
reflecting how distinct multipartite entanglement geometries redistribute
extractable work under the atmospheric thermal gradient.

\begin{table*}[tb]
\centering
\setlength{\tabcolsep}{4pt}
\begin{tabular}{ccccccc}
\toprule
$D_{HH}$ &
$r/r_h$ &
$\mathcal{E}^{\min}\!\left(\rho^{\mathrm{GHZ}}_{AB_{\mathrm I}C_{\mathrm I}}\right)$ &
$\mathcal{E}^{\min}\!\left(\rho^{\mathrm{GHZ}}_{AB_{\mathrm I}C_{\mathrm{II}}}\right)$ &
$\mathcal{E}^{\max}\!\left(\rho^{\mathrm{W}}_{AB_{\mathrm I}C_{\mathrm I}}\right)$ &
$\mathcal{E}^{\min}\!\left(\rho^{\mathrm{W}}_{AB_{\mathrm{II}}C_{\mathrm{II}}}\right)$ &
$\mathcal{E}^{\max}\!\left(\rho^{\mathrm{W}}_{AB_{\mathrm I}C_{\mathrm{II}}}\right)$ \\
\midrule
23.03 & 1.433 & 0.9372 & 1.9372 & 0.0851 & 1.3319 & 0.7085 \\
50.00 & 1.467 & 0.8355 & 1.8355 & 0.2302 & 1.3224 & 0.7763 \\
75.00 & 1.477 & 0.7712 & 1.7712 & 0.3282 & 1.3102 & 0.8192 \\
100.0 & 1.483 & 0.7265 & 1.7265 & 0.4002 & 1.2977 & 0.8490 \\
\bottomrule
\end{tabular}
\caption{Ergotropy extrema of the GHZ and W states in the black-hole
quantum atmosphere for different $D_{HH}$ (fixed $\alpha=1/\sqrt{2}$
and $\omega=1$). After the common extremal position
$r/r_h$, the columns list, in order: the minima of
$\mathcal{E}(\rho^{\scriptscriptstyle\mathrm{GHZ}}_{AB_{\mathrm I}C_{\mathrm I}})$
and
$\mathcal{E}(\rho^{\scriptscriptstyle\mathrm{GHZ}}_{AB_{\mathrm I}C_{\mathrm{II}}})$,
followed by the maximum of
$\mathcal{E}(\rho^{\scriptscriptstyle\mathrm{W}}_{AB_{\mathrm I}C_{\mathrm I}})$,
the minimum of
$\mathcal{E}(\rho^{\scriptscriptstyle\mathrm{W}}_{AB_{\mathrm{II}}C_{\mathrm{II}}})$,
and the maximum of
$\mathcal{E}(\rho^{\scriptscriptstyle\mathrm{W}}_{AB_{\mathrm I}C_{\mathrm{II}}})$.}
\label{tab:extrema_compare}
\end{table*}


\section{DISCUSSION AND CONCLUSIONS}\label{conclusion}
In this work, based on the tripartite GHZ and W entangled states shared by Alice, Bob, and Charlie, 
we study the variation of ergotropy under uniform proper acceleration near the Schwarzschild black hole within the Hartle–Hawking vacuum. 
Specifically, Alice stays in the inertial asymptotically flat region, whereas Bob and Charlie are subject to gravitational thermal effects close to the event horizon. By retaining one of the two complementary modes (Region~$I$ or Region~$II$) for each of Bob and Charlie and tracing out the other, four independent reduced subsystems $\rho_{AB_\mathrm{I}C_\mathrm{I}}$, $\rho_{AB_\mathrm{I}C_\mathrm{II}}$, $\rho_{AB_\mathrm{II}C_\mathrm{I}}$, and $\rho_{AB_\mathrm{II}C_\mathrm{II}}$ are constructed for subsequent analysis.
The spatially dependent local Hartle–Hawking temperature is governed by the constant $D_{HH}$, which regulates the thermal parameters $\xi=\cos \zeta$, $\eta=\sin \zeta$ and further reshapes the radial distribution of extractable work for different normalized distances $r/r_h$ outside the event horizon.

Concretely, for the GHZ state, the magnitude of ergotropy is directly determined by the acceleration parameters, radial coordinate and subsystem partition scheme. 
Under symmetric acceleration $\zeta_b=\zeta_c=\zeta$, ergotropy degeneracy emerges such that $\mathcal{E}(\rho_{AB_\mathrm{I}C_\mathrm{I}})=\mathcal{E}(\rho_{AB_\mathrm{II}C_\mathrm{II}})$ and $\mathcal{E}(\rho_{AB_\mathrm{I}C_\mathrm{II}})=\mathcal{E}(\rho_{AB_\mathrm{II}C_\mathrm{I}})$, 
and a constant energy gap exists between these two groups of degenerate ergotropy values for arbitrary $r$ and $D_{HH}$. 
With the growth of acceleration or the reduction of radial distance, the extractable work decays monotonically, while a larger value of $D_{HH}$ slows down the rate at which ergotropy recovers to the asymptotic Hawking limit at large $r/r_h$.

Remarkably, the W state exhibits totally different thermodynamic behaviors relative to the GHZ state. A strict linear relation $\mathcal{E}(\rho_{AB_\mathrm{I}C_\mathrm{II}})=\tfrac12\left[\mathcal{E}(\rho_{AB_\mathrm{I}C_\mathrm{I}})+\mathcal{E}(\rho_{AB_\mathrm{II}C_\mathrm{II}})\right]$ holds for all parameter ranges, a feature that cannot be observed in the GHZ configuration. Such disparate evolution features stem from the distinct multipartite entanglement structures after eliminating the unobservable interior modes confined within the event horizon.

By characterizing the competing influences of radial distance and Hartle–Hawking constant $D_{HH}$ on ergotropy, we conclude that relativistic thermal corrections do not alter the intrinsic upper energy limit of quantum entanglement, yet ergotropy, as the quantitative metric of retrievable quantum work, is highly sensitive to gravitational decoherence, radial thermal gradients and entanglement redistribution induced by Schwarzschild spacetime. Furthermore, our results quantitatively reveal how thermal effects from the Hartle–Hawking vacuum modulate the maximal extractable work of tripartite entangled states, and clarify the fundamental distinction between the frame-invariant entanglement energy bound and the position-dependent ergotropy in curved-spacetime quantum thermodynamics.

    
\section*{ACKNOWLEDGMENTS}
This work is supported by the Natural Science Foundation of Hainan Province under Grant No. 125RC744; 
the China Scholarship Council (CSC).	
	
\nocite{*}
\bibliography{apssamp}
	
\appendix
	
\section{The reduced density matrices and eigenvalues of the GHZ-type tripartite state} \label{app_GHZ}
First, Alice stays in the inertial asymptotically flat region, whereas Bob and Charlie are subject to gravitational thermal effects close to the event horizon. By retaining one of the two complementary modes (Region~I or Region~II) for each of Bob and Charlie and tracing out the other, four independent reduced subsystems $\rho^{\scriptscriptstyle\mathrm{GHZ}}_{AB_\mathrm{I}C_\mathrm{I}}$, $\rho^{\scriptscriptstyle\mathrm{GHZ}}_{AB_\mathrm{I}C_\mathrm{II}}$, $\rho^{\scriptscriptstyle\mathrm{GHZ}}_{AB_\mathrm{II}C_\mathrm{I}}$, and $\rho^{\scriptscriptstyle\mathrm{GHZ}}_{AB_\mathrm{II}C_\mathrm{II}}$ are constructed for subsequent analysis.
Since the ergotropy is fully determined by the eigenvalue spectrum of the density matrix, we list below the explicit form of each reduced state followed by its diagonalization result.
\subsubsection{Reduced state~$\rho^{ \scriptscriptstyle\mathrm{GHZ}}_{AB_{\mathrm{I}}C_{\mathrm{I}}}$}
\begin{equation}
\rho^{\scriptscriptstyle\mathrm{GHZ}}_{AB_{\mathrm{I}} C_{\mathrm{I}}}=
\begin{pmatrix}
a_{1} & 0 & 0 & 0 & 0 & 0 & 0 & z_{1} \\
0 & a_{2} & 0 & 0 & 0 & 0 & 0 & 0 \\
0 & 0 & a_{3} & 0 & 0 & 0 & 0 & 0 \\
0 & 0 & 0 & a_{4} & 0 & 0 & 0 & 0 \\
0 & 0 & 0 & 0 & 0 & 0 & 0 & 0 \\
0 & 0 & 0 & 0 & 0 & 0 & 0 & 0 \\
0 & 0 & 0 & 0 & 0 & 0 & 0 & 0 \\
z_{1} & 0 & 0 & 0 & 0 & 0 & 0 & b_{1}
\end{pmatrix},
\end{equation}

where the matrix elements are given by
\begin{equation}
\begin{aligned}
b_{1}&=1-\alpha^{2}, 
z_{1}=\alpha\sqrt{1-\alpha^{2}}\cos \zeta_{b}\cos \zeta_{c},\\
a_{1}&=\alpha^{2}\cos^{2}\zeta_{b}\cos^{2}\zeta_{c}, 
a_{2}=\alpha^{2}\cos^{2}\zeta_{b}\sin^{2}\zeta_{c}, \\
a_{3}&=\alpha^{2}\sin^{2}\zeta_{b}\cos^{2}\zeta_{c}, 
a_{4}=\alpha^{2}\sin^{2}\zeta_{b}\sin^{2}\zeta_{c}.
\end{aligned}
\end{equation}

the eigenvalues are as follows:
\begin{equation}
\begin{aligned}
\lambda_{0}&=\lambda_{1}=\lambda_{2}=\lambda_{3}=0,\\
\lambda_{4}&=\alpha^2\cos^2 \zeta_b\sin^2 \zeta_c,\\
\lambda_{5}&=\alpha^2\sin^2 \zeta_b\cos^2 \zeta_c,\\
\lambda_{6}&=\alpha^2\sin^2 \zeta_b\sin^2 \zeta_c,\\
\lambda_{7}&=1-\alpha^2\left(1-\cos^2\zeta_b\cos^2\zeta_c\right).
\end{aligned}
\end{equation}

\subsubsection{Reduced state~$\rho^{ \scriptscriptstyle\mathrm{GHZ}}_{AB_{\mathrm{I}}C_{\mathrm{II}}}$}
\begin{equation}
\rho^{\scriptscriptstyle\mathrm{GHZ}}_{AB_{\mathrm{I}} C_{\mathrm{II}}}=
\begin{pmatrix}
a_1 & 0 & 0 & 0 & 0 & 0 & 0 & 0 \\
0 & a_2 & 0 & 0 & 0 & 0 & z_2 & 0 \\
0 & 0 & a_3 & 0 & 0 & 0 & 0 & 0 \\
0 & 0 & 0 & a_4 & 0 & 0 & 0 & 0 \\
0 & 0 & 0 & 0 & 0 & 0 & 0 & 0 \\
0 & 0 & 0 & 0 & 0 & 0 & 0 & 0 \\
0 & z_2 & 0 & 0 & 0 & 0 & b_2 & 0 \\
0 & 0 & 0 & 0 & 0 & 0 & 0 & 0
\end{pmatrix}
\end{equation}
where the matrix elements are given by
\begin{equation}
\begin{aligned}
 b_2 &= 1-\alpha^2,\quad z_2=\alpha\sqrt{1-\alpha^2}\cos \zeta_b\sin \zeta_c,\\
a_1 &= \alpha^2\cos^2 \zeta_b\cos^2 \zeta_c,\quad a_2=\alpha^2\cos^2 \zeta_b\sin^2 \zeta_c,\\
a_3 &= \alpha^2\sin^2 \zeta_b\cos^2 \zeta_c,\quad a_4=\alpha^2\sin^2 \zeta_b\sin^2 \zeta_c.\\
\end{aligned}
\end{equation}
	
the eigenvalues are as follows:
\begin{equation}
\begin{aligned}
\lambda_0&=\lambda_1=\lambda_2=\lambda_3=0,\\
\lambda_4&=\alpha^2\cos^2\zeta_b\cos^2\zeta_c,\\
\lambda_5&=\alpha^2\sin^2\zeta_b\cos^2\zeta_c,\\
\lambda_6&=\alpha^2\sin^2\zeta_b\sin^2\zeta_c,\\
\lambda_7
&=1-\alpha^2\left(1-\cos^2\zeta_b\sin^2\zeta_c\right).
\end{aligned}
\end{equation}

\subsubsection{Reduced state~$\rho^{ \scriptscriptstyle\mathrm{GHZ}}_{AB_{\mathrm{II}}C_{\mathrm{I}}}$}
    
\begin{equation}
\rho^{\scriptscriptstyle\mathrm{GHZ}}_{AB_{\mathrm{II}}C_{\mathrm{I}}}=
\begin{pmatrix}
a_1 & 0 & 0 & 0 & 0 & 0 & 0 & 0 \\
0 & a_2 & 0 & 0 & 0 & 0 & 0 & 0 \\
0 & 0 & a_3 & 0 & 0 & z & 0 & 0 \\
0 & 0 & 0 & a_4 & 0 & 0 & 0 & 0 \\
0 & 0 & 0 & 0 & 0 & 0 & 0 & 0 \\
0 & 0 & z & 0 & 0 & b_2 & 0 & 0 \\
0 & 0 & 0 & 0 & 0 & 0 & 0 & 0 \\
0 & 0 & 0 & 0 & 0 & 0 & 0 & 0
\end{pmatrix},
\end{equation}
	
\begin{equation}
\begin{aligned}
b_2&=1-\alpha^2,\quad z=\alpha\sqrt{1-\alpha^2}\sin \zeta_b\cos \zeta_c,\\
a_1&=\alpha^2\cos^2\zeta_b\cos^2\zeta_c,\quad a_2=\alpha^2\cos^2\zeta_b\sin^2\zeta_c,\\
a_3&=\alpha^2\sin^2\zeta_b\cos^2\zeta_c,\quad a_4=\alpha^2\sin^2\zeta_b\sin^2\zeta_c.\\
\end{aligned}
\end{equation}
	
the eigenvalues are as follows:
\begin{equation}
\begin{aligned}
\lambda_0 &= \lambda_1 = \lambda_2 = \lambda_3 = 0,\\
\lambda_4 &=\alpha^2\cos^2 \zeta_b\cos^2 \zeta_c,\\
\lambda_5 &=\alpha^2\cos^2 \zeta_b\sin^2 \zeta_c,\\
\lambda_6 &=\alpha^2\sin^2 \zeta_b\sin^2 \zeta_c,\\
\lambda_7
&=1-\alpha^2\left(1-\sin^2\zeta_b\cos^2\zeta_c\right).
\end{aligned}
\end{equation}
    
\subsubsection{Reduced state~$\rho^{ \scriptscriptstyle\mathrm{GHZ}}_{AB_{\mathrm{II}}C_{\mathrm{II}}}$}
    
\begin{equation}
\rho^{\scriptscriptstyle\mathrm{GHZ}}_{AB_{\mathrm{II}} C_{\mathrm{II}}}=
\begin{pmatrix}
a_1 & 0 & 0 & 0 & 0 & 0 & 0 & 0 \\
0 & a_2 & 0 & 0 & 0 & 0 & 0 & 0 \\
0 & 0 & a_3 & 0 & 0 & 0 & 0 & 0 \\
0 & 0 & 0 & a_4 & z_4 & 0 & 0 & 0 \\
0 & 0 & 0 & z_4 & b_4 & 0 & 0 & 0 \\
0 & 0 & 0 & 0 & 0 & 0 & 0 & 0 \\
0 & 0 & 0 & 0 & 0 & 0 & 0 & 0 \\
0 & 0 & 0 & 0 & 0 & 0 & 0 & 0
\end{pmatrix},
\end{equation}
where the matrix elements are given by
\begin{equation}
\begin{aligned}
\quad b_{4}&=1-\alpha^{2}, \quad z_{4}=\alpha \sqrt{1-\alpha^{2}} \sin \zeta_{b}\sin \zeta_{c},\\
a_{1}& =\alpha^{2} \cos^{2}\zeta_{b} \cos^{2}\zeta_{c}, 
a_{2} =\alpha^{2} \cos^{2}\zeta_{b} \sin^{2}\zeta_{c}, \\
a_{3}& =\alpha^{2} \sin^{2}\zeta_{b} \cos^{2}\zeta_{c}, 
a_{4} =\alpha^{2} \sin^{2}\zeta_{b} \sin^{2}\zeta_{c}.
\end{aligned}
\end{equation}
	
the eigenvalues are as follows:
\begin{equation}
\begin{aligned}
\lambda_0 &=\lambda_1 = \lambda_2 = \lambda_3 = 0,\\
\lambda_4 &=\alpha^2\cos^2 \zeta_b\cos^2 \zeta_c,\\
\lambda_5 &=\alpha^2\cos^2 \zeta_b\sin^2 \zeta_c,\\
\lambda_6 &=\alpha^2\sin^2 \zeta_b\cos^2 \zeta_c,\\
\lambda_7
&=1-\alpha^2\left(1-\sin^2\zeta_b\sin^2\zeta_c\right).
\end{aligned}
\end{equation}

Ordering each spectrum in ascending order and pairing it with the four-fold degenerate levels $\epsilon_{0,1,2,3}=-1$, $\epsilon_{4,5,6,7}=+1$ of $H =\sigma_z^{A}\sigma_z^{B}\sigma_z^{C}$ through Eq.~\ref{eq:ergotropy_def}, we arrive the ergotropy expression.
    
\section{The reduced density matrices and eigenvalues of the W-type tripartite state}\label{app_W}

Along the same lines as Appendix~A, by tracing over the complementary mode (Region~I or Region~II) of each of Bob and Charlie in the five-partite W state, with $\xi=\cos \zeta$, $\eta=\sin \zeta$ and $\xi^2+\eta^2=1$, one obtains the four reduced density operators and their spectra listed below, corresponding to the four combinations of the retained modes.
\subsubsection{Reduced state~$\rho^{ \scriptscriptstyle\mathrm{W}}_{AB_{\mathrm{I}}C_{\mathrm{I}}}$}

{\footnotesize\setlength{\arraycolsep}{3pt}\renewcommand{\arraystretch}{0.85}
\[\rho^{\scriptscriptstyle\mathrm{W}}_{AB_{\mathrm I} C_{\mathrm I}}=\frac{1}{3}
\begin{pmatrix}
0 & 0 & 0 & 0 & 0 & 0 & 0 & 0 \\
0 & \xi^2 & \xi^2 & 0 & \xi^3 & 0 & 0 & 0 \\
0 & \xi^2 & \xi^2 & 0 & \xi^3 & 0 & 0 & 0 \\
0 & 0 & 0 & 2\eta^2 & 0 & \xi\eta^2 & \xi\eta^2 & 0 \\
0 & \xi^3 & \xi^3 & 0 & \xi^4 & 0 & 0 & 0 \\
0 & 0 & 0 & \xi\eta^2 & 0 & \xi^2\eta^2 & 0 & 0 \\
0 & 0 & 0 & \xi\eta^2 & 0 & 0 & \xi^2\eta^2 & 0 \\
0 & 0 & 0 & 0 & 0 & 0 & 0 & \eta^4
\end{pmatrix}.
\]
}
Its eigenvalues are
\begin{align*}
& \lambda_1=\lambda_2=\lambda_3=\lambda_4 = 0, \\
& \lambda_5 = \frac{1}{3}\xi^2\big(3-\eta^2\big),\quad
\lambda_6 = \frac{1}{3}\xi^2\eta^2, \\
& \lambda_7 = \frac{1}{3}\eta^2\big(2+\xi^2\big),\quad
 \lambda_8 = \frac{1}{3}\eta^4.
\end{align*}
\subsubsection{Reduced state~$\rho^{\scriptscriptstyle\mathrm{W}}_{AB_{\mathrm I}C_{\mathrm{II}}}$ and
$\rho^{\scriptscriptstyle\mathrm{W}}_{AB_{\mathrm{II}}C_{\mathrm I}}$}

{\footnotesize\setlength{\arraycolsep}{3pt}\renewcommand{\arraystretch}{0.85}
\[\rho^{\scriptscriptstyle\mathrm{W}}_{AB_{\mathrm I} C_{\mathrm{II}}}=\rho^{\scriptscriptstyle\mathrm{W}}_{AB_{\mathrm{II}} C_{\mathrm I}}=\frac{1}{3}
\begin{pmatrix}
\xi^2 & 0 & 0 & \xi\eta & 0 & \xi^2\eta & 0 & 0 \\
0 & 0 & 0 & 0 & 0 & 0 & 0 & 0 \\
0 & 0 & 1 & 0 & \xi^3 & 0 & 0 & \eta^3 \\
\xi\eta & 0 & 0 & \eta^2 & 0 & \xi\eta^2 & 0 & 0 \\
0 & \xi^3 & 0 & \xi^4 & 0 & 0 & 0 & 0 \\
\xi^2\eta & 0 & 0 & \xi\eta^2 & 0 & \xi^2\eta^2 & 0 & 0 \\
0 & 0 & 0 & 0 & 0 & 0 & \xi^2\eta^2 & 0 \\
0 & 0 & \eta^3 & 0 & 0 & 0 & 0 & \eta^4
\end{pmatrix}.
\]
}
Its eigenvalues are
\begin{align*}
& \lambda_1=\lambda_2=\lambda_3=\lambda_4 = 0, \\
& \lambda_5 = \frac{\eta^2 \xi^2}{3}, \lambda_6 = \frac{\eta^2\xi^2}{3} + \frac{1}{3},\\
& \lambda_7 = \frac{\eta^4}{6} - \frac{1}{6}\sqrt{\eta^8 + 4\eta^6 - 2\eta^4 + 1} + \frac{1}{6}, \\
& \lambda_8 = \frac{\eta^4}{6} + \frac{1}{6}\sqrt{\eta^8 + 4\eta^6 - 2\eta^4 + 1} + \frac{1}{6}. \\
\end{align*}

\subsubsection{Reduced state~$\rho^{\scriptscriptstyle\mathrm{W}}_{AB_{\mathrm{II}}C_{\mathrm{II}}}$}

{\footnotesize\setlength{\arraycolsep}{3pt}\renewcommand{\arraystretch}{0.85}
\[\rho^{\scriptscriptstyle\mathrm{W}}_{AB_{\mathrm{II}} C_{\mathrm{II}}}=\frac{1}{3}
\begin{pmatrix}
2\xi^2 & 0 & 0 & 0 & 0 & \xi^2\eta & \xi^2\eta & 0 \\
0 & \eta^2 & \eta^2 & 0 & 0 & 0 & 0 & \eta^3 \\
0 & \eta^2 & \eta^2 & 0 & 0 & 0 & 0 & \eta^3 \\
0 & 0 & 0 & 0 & 0 & 0 & 0 & 0 \\
0 & 0 & 0 & 0 & \xi^4 & 0 & 0 & 0 \\
\xi^2\eta & 0 & 0 & 0 & 0 & \xi^2\eta^2 & 0 & 0 \\
\xi^2\eta & 0 & 0 & 0 & 0 & 0 & \xi^2\eta^2 & 0 \\
0 & \eta^3 & \eta^3 & 0 & 0 & 0 & 0 & \eta^4
\end{pmatrix}.
\]
}
Its eigenvalues are
\begin{align*}
& \lambda_1=\lambda_2=\lambda_3=\lambda_4 = 0, \\
& \lambda_5 = \frac{\xi^4}{3},
\lambda_6 = \frac{\eta^2\xi^2}{3},\\
& \lambda_7 = \frac{\eta^4}{3} + \frac{2\eta^2}{3},
\lambda_8 = \frac{\eta^2\xi^2}{3} + \frac{2\xi^2}{3}. \\
\end{align*}

Combining these spectra with the same Hamiltonian
$H =\sigma_z^{A}\sigma_z^{B}\sigma_z^{C}$, 
whose two energy levels are each
four-fold degenerate, yields the W state ergotropies.

\end{document}